\documentstyle[11pt,newpasp,twoside,epsf]{article}
\begin{document}

\def\etal{{et~al.\,}}
\def\mic{$\mu$m\ }

\title{Modeling Brown Dwarfs, L Dwarfs, and T Dwarfs}
\author{Adam Burrows}
\affil{Department of Astronomy, The University of Arizona, Tucson, AZ, USA 85721}

\begin{abstract}

In this contribution, I touch on a subset of our recent efforts in spectral 
and opacity modeling aimed at improving our understanding of  
brown dwarfs, L dwarfs, and T dwarfs.  I discuss theoretical 
calculations of the alkali line profiles, newly generated CrH
opacities, new evidence for refractory rainout 
in T dwarfs from optical spectral measurements,
and the distinction between brown dwarfs and planets.

\end{abstract}

\section{Introduction}

The subject of brown dwarfs (and substellar-mass objects in general) 
is entering a new phase of rapid expansion and discovery.  More than 200
L dwarfs are now known, and they are joined by $\sim$30 T dwarfs.  This development 
requires a corresponding expansion in theoretical effort, involving at its core 
evolutionary, spectral, and compositional modeling.  Crucial to spectral modeling
are molecular and atomic opacities, many of which have not been addressed before
with the degree of seriousness and completeness that standard stellar atmospheres
studies have long enjoyed.  However, the pace of relevant spectral and opacity 
calculations is accelerating and in this spirit I summarize in this paper
a few such topics of recent interest.  In \S\ref{rain}, I show new proof of the ``rainout"
and settling of refractory elements in T dwarfs.  Rainout leaves 
as one of its consequences the lower-temperature reaches of T dwarf
atmospheres enhanced in sodium and potassium atoms.  In \S\ref{profile}, I touch on  
some new calculations of the alkali line wings to many 1000's of \AA\ detunings.
The red wing of the K I line centered at 0.77 \mic in particular defines 
much of the T dwarf continuum between 0.77 \mic and 1.0 \mic.
This is followed in \S\ref{crh} by a short discussion of the new CrH opacities recently 
generated by our group.  Finally, and a bit tentatively, in \S\ref{name} I finish with a short
discussion on nomenclature, a subject that continues to exercise the brown dwarf and 
extrasolar planet communities.

\section{A New Proof of Rainout/Depletion in T Dwarfs}
\label{rain}

\begin{figure}
\plotone{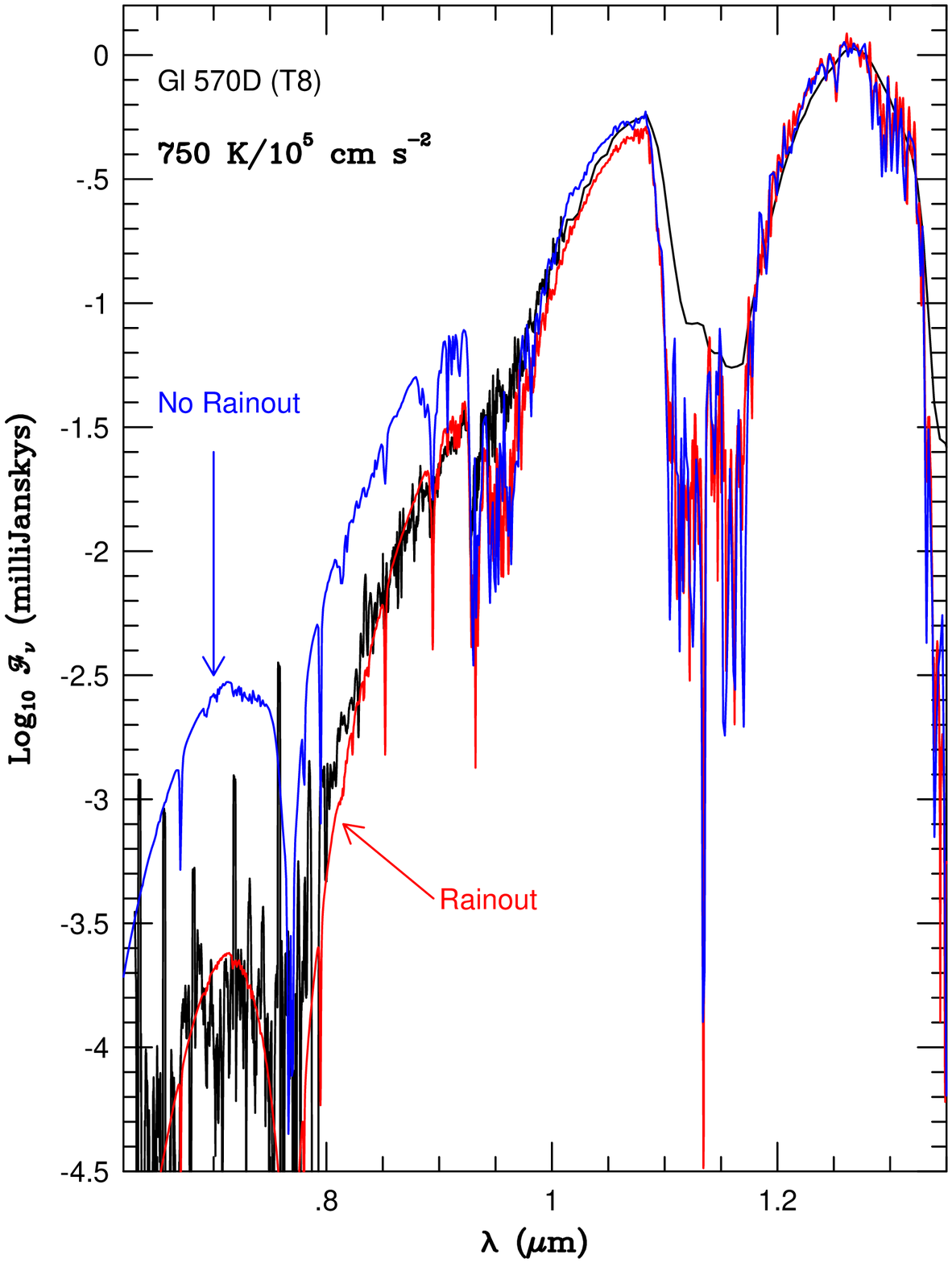}
\caption{A comparison of the measured absolute flux (F$_{\nu}$), in milliJanskys) of Gliese 570D (thick, black)
with two solar-metallicity model spectra for wavelengths ($\lambda$) from 0.6 $\mu$m to 1.4 $\mu$m.
The best-fit line incorporates the rainout of silicates and the resulting persistence
of Na and K at altitude, while the upper line is for strict chemical equilibrium without rainout.
From the significant ($\times 10$) difference between the data and the ``No rainout" model
near 0.7 $\mu$m, we see evidence for the role of rainout in T dwarf atmospheres.
The Gliese 570D models assume T/gravity = [750 K/10$^{5}$ cm s$^{-2}$].
\label{fig:1}}
\end{figure}

As explained in 
Burrows, Marley, and Sharp (2000, BMS), rainout depletes the atmospheres
of the refractory elements Ca, Al, Mg, Fe, and Si.  
In particular, rainout suppresses the formation of alkali feldspars and
enables atomic Na and K to survive to lower temperatures and pressures, at which
point they form Na$_2$S(c) and KCl(c) (Lodders 1999).
As a consequence, their influence on the emergent spectrum in the ``optical" between 0.5 \mic
and 1.1 \mic is enhanced.  This effect is seen dramatically for the first time in Fig. \ref{fig:1}.
At $\sim 0.7 \mu{\rm m}$, the data for Gliese 570D (Burgasser et al 2001), the coolest known T dwarf,
clearly favor the ``rainout" model.  This fact reiterates the conclusion of BMS that the wings
of the Na D and K I (7700\AA) resonance lines and aggressive rainout of heavy metals
(with the resulting enhancement of the sodium and potassium abundances at altitude)
are required to fit the T dwarf data shortward of 1.0 \mic.
It is shortward of $\sim$1.0 $\mu{\rm m}$ that the alkali chemistry
can be most readily probed (Burrows et al. 2002a).

\section{Alkali Metal Line Profiles}
\label{profile}

To continue on the alkali theme, in standard stellar atmospheres, atomic lines are superposed on
a background continuum and the concepts of equivalent width and
curve of growth make conceptual and practical sense.  An individual
line is but a perturbation on the local spectrum.  However, due to
the rainout of metals in cool atmospheres and the consequent
paucity of continuum and alternate opacity sources
between 0.4 \mic and 1.0 \mic, the wings of the strong resonance doublets
centered at $\sim$7700 \AA\ (K I) and 5890 \AA (Na-D)
assume the role of the continuum throughout most of this
spectral range (BMS; Tsuji, Ohnaka, and Aoki 1999).
In particular, the red wing of
the $4s^2S_{1/2}-4p^2P_{3/2}$ transitions of K I provides
the pseudo-continuum in cool molecular atmospheres
all the way from 0.77 \mic to $\sim$1.0 \mic.  The Na-D doublet, centered
as it is in the middle of the visible, determines the true color of brown dwarfs
(magenta/purple; Burrows et al. 2001).

Hence, whereas in traditional stellar atmospheres the
Lorentzian core and Gaussian wings of a line are not generally of relevance beyond
$\sim$20 \AA\ detunings ($\Delta\lambda$ from the line core), in cool substellar
atmospheres the relevant reach of the Na I and K I resonance lines can be
thousands of \AA.  Given this, to achieve accurate spectral fits
for brown dwarf, L dwarf, T dwarf, and hot giant planet atmospheres,
the shapes of the far wings of these alkali lines as a function of
pressure and temperature must be ascertained.

\begin{figure}
\plotone{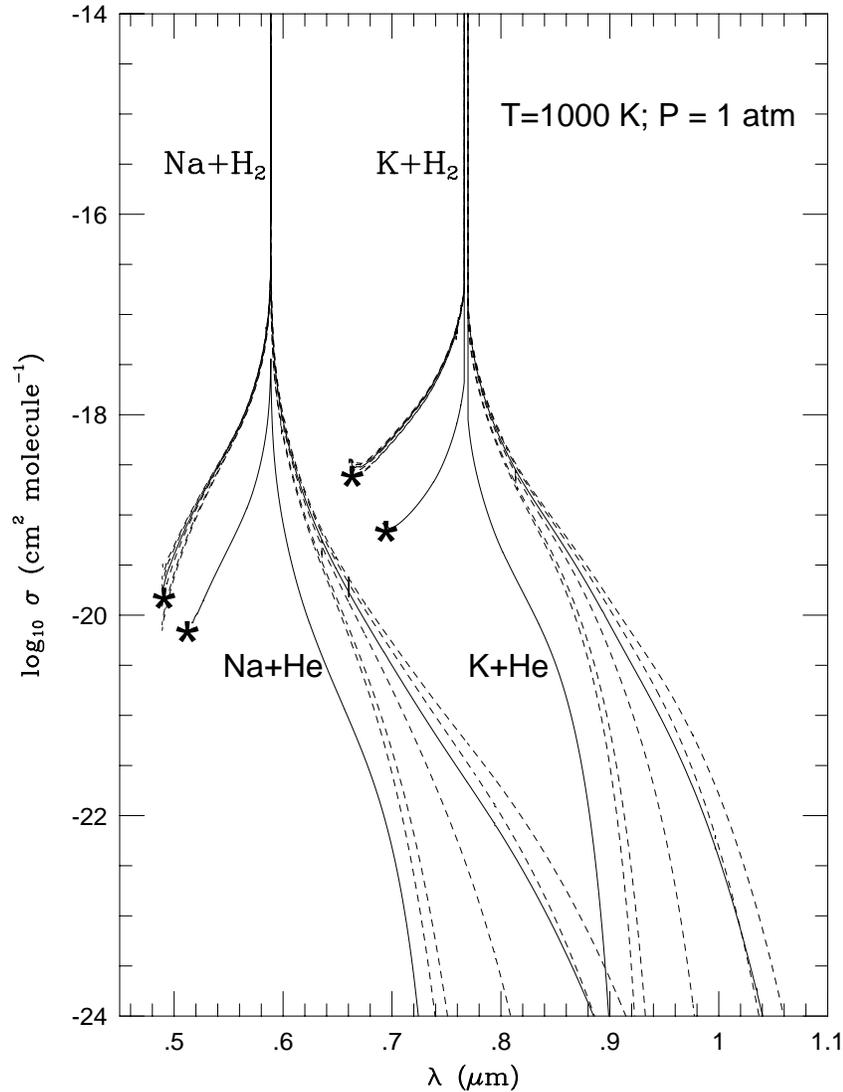}
\caption{Absorption cross sections (in cm$^2$) versus wavelength (in \mic) for the
Na-D and K I doublets at 5890 \AA\ and 7700 \AA, respectively, using the quasi-static
theory of wing line profiles.  The Lorentzian line cores are not included on this plot.  Cross sections
are shown for the different orientations of the H$_2$ molecule
($\theta = 0^{\circ}, 20^{\circ}, 45^{\circ}, 70^{\circ}, 90^{\circ}$), as well as after
integrating over angle (solid).   The behavior is monotonic with angle, from steepest (small angle)
to shallowest (large angle).  Also depicted are the cross sections due to
perturbations by the spherical helium atom (with C$_{\infty}$ symmetry).
The asterisks indicate the positions of the rainbow satellites on the blue wings.
A temperature of 1000 K and a total pressure of one atmosphere have been used.
The partial pressure of He is assumed to be $\sim$10\% of this total (Anders and Grevesse 1989).
\label{fig:2}}
\end{figure}

Recently, Burrows and Volobuyev (2002c) performed ab initio calculations of the energy shifts
of the ground and excited states of sodium and potassium
immersed in H$_2$- and helium-rich
atmospheres and have obtained the opacity profiles of the red and blue wings
of the Na-D and K I resonance lines using the Unified Franck-Condon formalism (Szudy and Baylis 1975,1996).
The absorption spectra as a function of photon wavelength for the D$_2$ and D$_1$
lines of Na and K centered near 5890 \AA\ and 7700 \AA,
respectively, are depicted in Fig. \ref{fig:2} at a temperature of 1000 K
and a total pressure of one atmosphere.  The contributions at
various orientation angles of H$_2$, the angle-integrated spectra (solid),
and the results for the Na+He and K+He systems are shown.  The Lorentzian cores
are not included on this plot.  

As seen in Fig. \ref{fig:2}, the cutoff on the red
wing of the potassium feature due to an exponential term containing the ground-state
interaction potential of the K+H$_2$ system is seen to be situated between 0.95 \mic
and 1.0 \mic, close to the 0.98 \mic used in Burrows et al. (2002c).  The cutoff of the
red wing of the Na feature is near 0.8 \mic, but is a bit more gradual.  The corresponding
cutoffs for the alkali-He systems are more abrupt, but of less importance due to the lower abundance of helium.
Importantly, unlike in the algorithm of BMS, there are no free parameters
for this theory of line profiles; the cutoff on the red wing of the K I feature
appears naturally.  Fig. \ref{fig:2} also makes clear that the red and 
blue wings are asymmetrical.

\section{New CrH Opacities}
\label{crh}

The 1-0 and 0-0 bands of the A$^6\Sigma^{+}$ -- X$^6\Sigma^{+}$
transition of CrH are used as primary markers for the
new L dwarf spectral class (Kirkpatrick et al. 1999a,b).
Hence, accurate CrH line lists and oscillator strengths
are needed to calculate the CrH opacities now used to model the spectral energy
distributions of these transitional and substellar objects.

\begin{figure}
\plotone{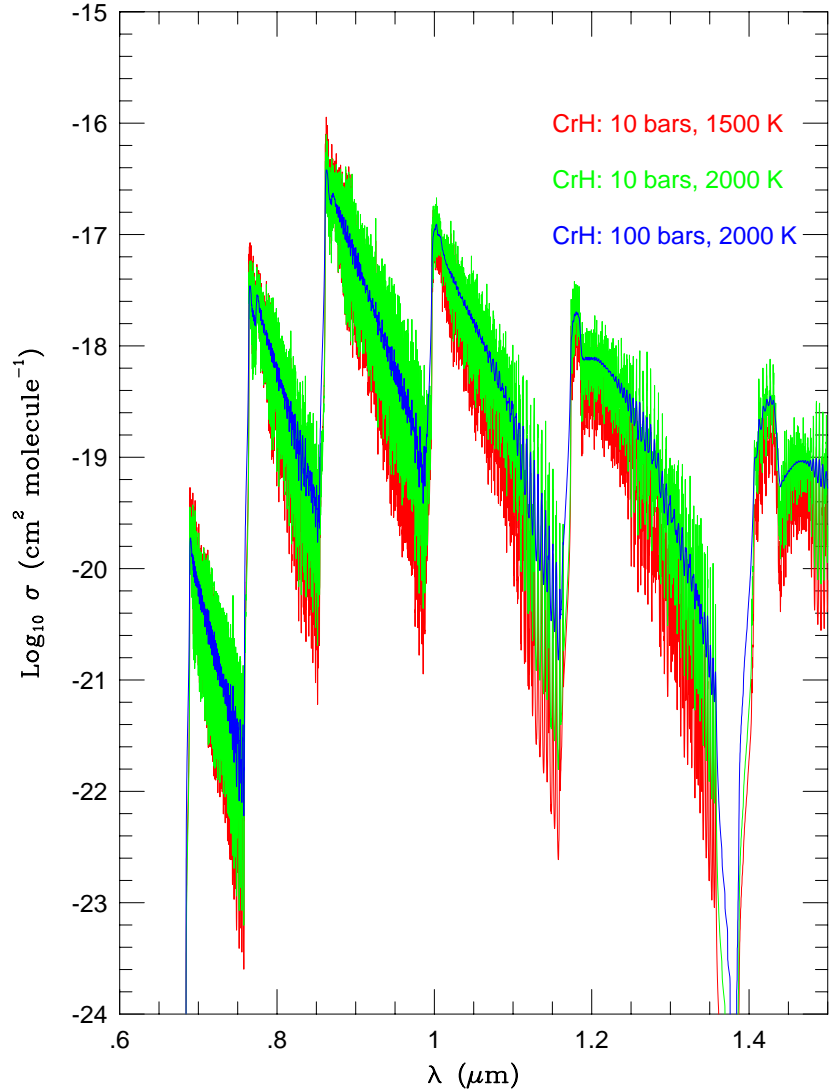}
\caption{The logarithm (base ten) of the absorption cross section of CrH versus
wavelength (in microns) from $\sim$0.7 $\mu$m
to 1.5 $\mu$m, at various temperatures and pressures.  The narrow (central) curve at 100
bars and 2000 K depicts the effect of large pressure broadening (when compared with the other curves
at 10 bars).  A comparison of the slightly upper curve (10 bars, 2000 K)
with the lower curve (10 bars, 1500 K) portrays the effect of increasing temperature.  The A--X 0-0 band is
the strongest and is the third from the left near $\sim$0.9 $\mu$m.
As the temperature and pressure decrease the cross section range in a given
band widens and manifests larger variation with wavelength.
\label{fig:3}}
\end{figure}

Burrows et al. (2002b) have recently created a program
for generating CrH cross sections for the six band sequences with band heads from
$\sim$0.7 \mic to $\sim$ 1.4 \mic.  
Figure \ref{fig:3} depicts the resulting CrH cross
sections for representative pressures
and temperatures at which CrH is typically found in substellar atmospheres.
Specifically, opacity spectra for $T/P$ pairs of 1500 K/10 bars, 2000
K/10 bars, and 2000 K/100 bars are portrayed.   The
corresponding opacities using the older database for the
0--0 transition of CrH of Ram, Jarman, and Bernath (1993), as calculated by R. Freedman (private communication),
are two orders of magnitude weaker.   The lion's share of the difference
between the old and the new CrH opacities can be traced to an increase by a factor of $\sim$13.5
in the oscillator strength and to a previously inappropriate division by 6,
the electronic spin degeneracy factor.  The A--X 0--0
band at $\sim$0.86 microns (just shortward of the
neighboring FeH feature at $\sim$0.87 microns) and the A--X 0--1 CrH band
near $\sim$0.997 microns (just longward of the classic Wing-Ford band of FeH)
are prominent and, in principle, diagnostic features in measured L dwarf
spectra (Kirkpatrick et al. 1999a,b).

\begin{figure}
\plotone{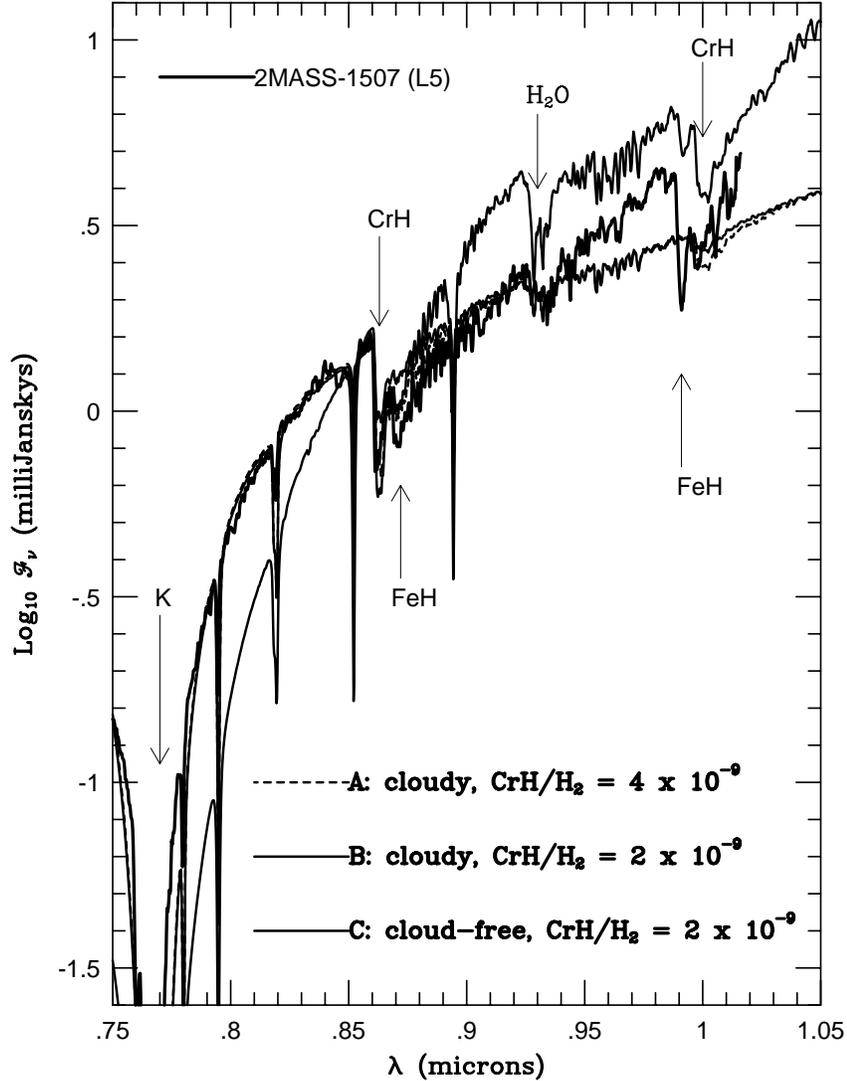}
\caption{The log (base ten) of the absolute flux density (${\cal F}_\nu$) in milliJanskys versus wavelength
($\lambda$) in microns from 0.75 $\mu$m to 1.05 $\mu$m for self-consistent
theoretical solar-metallicity models of the L5 dwarf 2MASS-1507.
Also included are the corresponding data (thick) for 2MASS-1507
from Kirkpatrick \etal (1999b).  All models are for T$_{\rm eff}$ = 1700 K and a gravity of $10^{5.5}$ cm s$^{-2}$.
The dashed line depicts a model (A) with a forsterite cloud and a CrH/H$_2$ number abundance ratio of $4\times 10^{-9}$,
the solid line (almost on top of the dashed line, except near CrH features) 
depicts a model (B) with a forsterite cloud and a CrH/H$_2$ number abundance ratio of $2\times 10^{-9}$,
and the steeper line depicts a cloud-free model (C) with a CrH/H$_2$ number abundance ratio of $2\times 10^{-9}$.
Below temperatures of 1400 K, the CrH abundance was set to zero.
Indicated with arrows are the positions of the CrH, FeH, H$_2$O, and K I (7700\AA) features in this
spectral range.  Also prominent are the Cs I lines at 8523 \AA\ and 8946 \AA, the Na I line at 8195 \AA,
and the Rb I lines at 7802 \AA\ and 7949 \AA.  These spectra
have been deresolved to an $R$($\lambda/{\Delta\lambda}$) of 1000.
\label{fig:4}}
\end{figure}

Since CrH is a defining molecule of the L dwarf spectroscopic class, accurate
opacities as a function of temperature and pressure are necessary for
spectral syntheses and to extract CrH abundances for L dwarf atmospheres.
Burrows et al. (2002b) have used the new theoretical opacity data to obtain 
an abundance for the L5 dwarf 2MASSI J1507038-151648 (Kirkpatrick et al. 1999b). 
Figure \ref{fig:4} depicts this comparison.  The CrH/H$_2$ number ratio
they find is $\sim 2-4\times 10^{-9}$, in reasonable agreement with expectations.

\section{What's in a Name?}
\label{name}

   The deuterium burning cut was used originally by some to distinguish planets from
brown dwarfs.  This was partly because of a figure in Burrows et al. (1997) that arbitrarily
distinguished evolutionary trajectories in luminosity/age space
on this basis, calling one class that burned tens of percent or more
of their deuterium brown dwarfs and the others that did not ``planets"
(as I recall, the original figure had the quotes as well).  Also, one
has the unfounded prejudice that a planet should not have a thermonuclear
phase.

   One should distinguish giant planets from brown dwarfs by their mode
of formation, by their origin.  Since an object's origin is not easily
``detected" or discerned, the situation is ready-made for confusion.
In fact, there is really no need to define and codify the nomenclature
nor the name to be applied to any particular object or class of objects
at this time; people just want (perhaps a bit too rigidly and with
unjustified Procrustean zeal) to name things, or to stuff objects
into cubbyholes.

   There is ambiguity in the provenance of these objects and this ambiguity
will be with us for a while.  If one uses a particular term (planet or brown
dwarf), one should always include the definition one is employing.
Not to do so implies that a standard, acceptable definition has been
arrived at; despite what some have written and the IAU, such is not
the case.

   Hence, formation history is a key, but not yet known. One would
expect a planet (or most of them) to be orbiting a star (or brown dwarf).
There may be metallicity (composition) differences between objects
with the same mass, but different origins or modes of formation.
Eccentricity is probably not a good discriminant (though this is not known).
Very few ``planets" are expected to be flung out of their cradles.
Brown dwarfs are just the tail of the stellar mass distribution,
but the mass distributions of brown dwarfs and giant planets may
overlap.

   When classifying a newly-discovered substellar-mass object,
one can use a variety of ``reasonable" criteria, but the tentative
classification is just that, tentative.  The usefulness of a classification
scheme at this stage in the development of the two related subjects of
brown dwarfs and giant planets is inversely proportional to its
rigidity.  A flexible and open-minded philosophy towards nomenclature
is best, which more data and information will progressively guide
towards a more and more reasonable naming scheme.  It may be that
the mass distributions of planets and brown dwarfs don't overlap, that
their compositions will enable one to distinguish one from the other,
etc. Until then, I prefer a bit of ambiguity to the illusion of clarity
and the pretense of certainty.  When there is fog, to acknowledge
its presence is more honest than to try to ignore it.

\acknowledgments

The author thanks Bill Hubbard, Jonathan Lunine, 
David Sudarsky, Ivan Hubeny, Christopher Sharp, Drew Milsom, 
Maxim Volobuyev, Curtis Cooper, Adam Burgasser, Davy Kirkpatrick, 
and Jonathan Fortney for informative exchanges and collaboration, as well as  
NASA for its financial support via 
grants NAG5-10760 and NAG5-10629.

% figure 1

% figure 2

% figure 3

% figure 4

\end{document}